\documentclass[%
 amsmath,amssymb,
 aps,
pra,
]{revtex4-2}

\usepackage{graphicx}
\usepackage{dcolumn}
\usepackage{bm}
\usepackage{lineno}
\usepackage{multirow}
\usepackage{bigstrut}
\usepackage{xcolor}

\usepackage{lineno}
\usepackage{multirow}
\usepackage{bigstrut}
\usepackage{xcolor}

\usepackage{color}
\usepackage{verbatim}
\usepackage{array}
\usepackage{epsfig}
\usepackage[ragged]{sidecap}
\usepackage{dsfont}
\usepackage{ulem}
\usepackage{mathrsfs}

\usepackage{caption}
\usepackage{subcaption}
\usepackage{gensymb}

\usepackage{siunitx}
\usepackage{url}
\usepackage{hyperref}
\usepackage{dcolumn}
\usepackage{textgreek}

\usepackage{soul,xcolor}

\newcommand{\etal}{{\it et al. }} 
\newcommand{\abi}{{\it ab initio }}

\begin{document}

\setstcolor{magenta}

\preprint{APS/123-QED}

%
%
%

\title{Experimental and theoretical approaches for determining the K-shell fluorescence yield of carbon}


%
%

\author{Philipp H\"onicke}
\affiliation{Physikalisch-Technische Bundesanstalt (PTB), Abbestr. 2-12, 10587 Berlin, Germany}
\email{philipp.hoenicke@ptb.de}
\author{Rainer Unterumsberger}
\affiliation{Physikalisch-Technische Bundesanstalt (PTB), Abbestr. 2-12, 10587 Berlin, Germany}
\author{Nils Wauschkuhn}
\affiliation{Physikalisch-Technische Bundesanstalt (PTB), Abbestr. 2-12, 10587 Berlin, Germany}
\author{Markus Kr\"amer}
\affiliation{AXO DRESDEN GmbH, Gasanstaltstr. 8b, 01237 Dresden, Germany}
\author{Burkhard Beckhoff}
\affiliation{Physikalisch-Technische Bundesanstalt (PTB), Abbestr. 2-12, 10587 Berlin, Germany}
\author{Paul Indelicato}
\affiliation{Laboratoire Kastler Brossel, Sorbonne Universit\'e, CNRS, ENS-PSL Research University, Coll\`ege de France, Case\ 74;\ 4, place Jussieu, F-75005 Paris, France}
\author{Jorge Sampaio}
\affiliation{LIP - Laboratory for Instrumentation and Experimental Particle Physics, Faculdade de Ci\^encias da Universidade de Lisboa, Campo Grande, C8, 1749-016 Lisboa, Portugal}
\author{Jos\'{e} Pires Marques}
\affiliation{LIP - Laboratory for Instrumentation and Experimental Particle Physics, Faculdade de Ci\^encias da Universidade de Lisboa, Campo Grande, C8, 1749-016 Lisboa, Portugal}
\author{Mauro Guerra}
\affiliation{Laboratory of Instrumentation, Biomedical Engineering and Radiation Physics (LIBPhys-UNL), Department of Physics, NOVA School of Science and Technology, NOVA University Lisbon, 2829-516 Caparica, Portugal}
\author{Fernando Parente}
\affiliation{Laboratory of Instrumentation, Biomedical Engineering and Radiation Physics (LIBPhys-UNL), Department of Physics, NOVA School of Science and Technology, NOVA University Lisbon, 2829-516 Caparica, Portugal}
\author{Jos\'{e} Paulo Santos}
\affiliation{Laboratory of Instrumentation, Biomedical Engineering and Radiation Physics (LIBPhys-UNL), Department of Physics, NOVA School of Science and Technology, NOVA University Lisbon, 2829-516 Caparica, Portugal}


%
%

\date{\today}

\begin{abstract}

The knowledge of atomic fundamental parameters, such as the fluorescence yields with low uncertainties, is of decisive importance in elemental
quantification involving X-ray fluorescence analysis techniques. However, especially for the low-Z elements, the available literature data are either of poor quality, of unknown or very large uncertainty, or both. For this reason, the K-shell  fluorescence yield of carbon was determined in the PTB laboratory at the synchrotron radiation facility BESSY II. In addition, theoretical calculations of the same parameter were performed using the multiconfiguration Dirac-Fock method, including relativistic and quantum electrodynamics (QED) corrections. Both values obtained in this work are compared to the corresponding available literature data.


\end{abstract}

\maketitle



\section{\label{sec:level1}Introduction}
The knowledge of accurate fluorescence yields in the X-ray regime is of great importance in many areas of physics and technology, such as fundamental physics~\cite{Schmidt:2018ge}, biological studies~\cite{Guimaraes:2012jw,Pessanha:2018ea}, biomedical ~\cite{Pessanha:2019du, Smuda2021}, environmental sciences~\cite{PeLeveSantos:2013et}, spectroscopy~\cite{Kasthurirangan:2014fk, Cara_2020}, plasma physics~\cite{Santos:2010dx}, nanoelectronics and microelectronics~\cite{Hiller2020,Hoenicke2020}, and astrophysics~\cite{Shah:2019dg}.

The majority of the available experimental, theoretical and empirical values of X-ray fluorescence (XRF) yields for different elements were obtained more than forty years ago. The calculations were mainly non-relativistic \cite{RMP38_513,PRA2_273,RMP44_716}, and the measurements were obtained with poor precision comparing with the modern technological possibilities and standards. Furthermore, the significant discrepancies observed between these datasets call for efforts to revisit and update fundamental parameter values with high precision measurements and state-of-art calculations. 

Especially for low-Z elements, e.g. carbon or oxygen, the status of the available data in the literature is not satisfying. This is due to the fact that there is only an estimated uncertainty budget available for low-Z element fluorescence yields which dates back to the 1970's \cite{Krause1979}. Furthermore, those estimated uncertainties are in the order of up to 40 \%, a value too large for most modern applications of e.g. quantitative X-ray fluorescence analysis. 

Carbon, as a very prominent element of the low-Z group, is ubiquitous in about all scientific areas, from biology and chemistry to technological applications. Due to its pronounced presence in nature, there is also a high interest in being able to reliably quantify carbon in quantitative XRF, for instance. Unfortunately, the state of available data on the carbon K-shell fluorescence yield is not satisfying due to the aforementioned reasons. Thus, we present new experimental and theoretical determinations of the carbon K-shell fluorescence yield employing free-standing thin foils of elemental carbon, polyimide, parylene, and silicon carbide. In addition, we have used a carbon thin film on silicon for which combined reference-free grazing incidence X-ray fluorescence (GIXRF) and X-ray reflectometry experiments have been performed \cite{Hoenicke2019}. The experimental determinations were carried out at the PTB laboratory at BESSY II, and the \abi theoretical value was obtained within the multiconfiguration Dirac-Fock (MCDF) framework.

%

\section{\label{sec:level2}Experimental determination}
\subsection{\label{sec:level21}Thin-foil experiments}
The work for the experimental determination of the fundamental parameters was carried out at the plane grating monochromator (PGM) beamline\cite{F.Senf1998} for undulator radiation at the BESSY II electron storage ring. This beamline is located in the PTB laboratory at BESSY II \cite{B.Beckhoff2009c} and provides soft X-ray radiation of high spectral purity in the photon energy range from 78 eV to 1860 eV. One advantage of this beamline is that it provides soft X-ray radiation at a high radiant power which is up to three orders of magnitude higher than for a typical bending magnet monochromator beamline. In addition, a slight detuning of the undulator harmonic energy against the PGM allows for an improved higher-order suppression capability in conjunction with the red shift of higher-order harmonics of the undulator. To further reduce the higher-order contributions, dedicated attenuation filters are used between the exit slit of the beamline and the focal plane. Depending on the operational parameters, stray light contributions of about 0.5 \% to 1 \% have to be taken into account. The uncertainty of the energy scale of the PGM is in the $10^{-4}$ range. For the calibration of the PGM energy scale typical resonance lines of Kr, Ar, and Ne gases are used\cite{FScholze2001}.

The experiments were carried out using two in-house developed ultrahigh vacuum chambers\cite{M.Kolbe2005a,J.Lubeck2013}. Both instruments are equipped with calibrated photo diodes and an energy-dispersive silicon drift detector (SDD) with experimentally determined response functions and radiometrically calibrated detection efficiency\cite{F.Scholze2009}. Each sample can be placed into the center of the respective chamber by means of an x-y scanning stage. The incident angle $\Psi_{in}$ between the surface of the sample and the incoming beam was set to 45$^\circ$ for all experiments excluding the GIXRF-XRR data.

For the experimental determination of the K-shell fluorescence yield of carbon, free standing thin foils of carbon (nominal thickness of 100 nm), polyimide (nominal thickness of 135 nm), parylene (nominal thickness of 100 nm), and silicon carbide (nominal thickness of 150 nm) were used.  Both fluorescence- and transmission experiments were conducted in the photon energy range around the carbon K-absorption edge. From the photon energy dependent transmission of each foil, sample specific mass attenuation factors $\mu_S(E_0)\rho d$ \cite{Unterumsberger2018} can be derived and used for the calculation of the attenuation correction factor $M_{i,E0}$ which is then independent from any database values for mass attenuation coefficients.

The sample specific fluorescence production yield $\sigma_K(E_0)\rho d$ for the K-shell of the respective chemical element at a given photon energy $E_0$ can be calculated according to Eq. \ref{eq:prodCS}, where $\omega_K$ is the K-shell fluorescence yield and $\tau_K(E_0)\rho d$ is the sample specific K-shell photoionization cross section for photons of energy $E_0$.

\begin{equation}
\sigma_K(E_0)\rho d = \omega_K \tau_K(E_0)\rho d = \frac{\Phi^d_i(E_0)M_{i,E0}}{\Phi_0(E_0)\frac{\Omega}{4\pi}}
\label{eq:prodCS}
\end{equation}
with
\begin{equation}
M_{i,E0} = \frac{(\frac{\mu_S(E_0)\rho d}{sin \theta_{in}}+\frac{\mu_S(E_i)\rho d}{sin \theta_{out}})}{(1-exp[-(\frac{\mu_S(E_0)\rho d}{sin \theta_{in}}+\frac{\mu_S(E_i)\rho d}{sin \theta_{out}})])}
\label{eq:M}
\end{equation}

The fluorescence photon flux $\Phi^d_i(E_0)$ is derived from the recorded fluorescence spectra by means of a deconvolution procedure. The detector response functions for all relevant fluorescence lines as well as relevant background contributions, e.g. bremsstrahlung, originating from photo-electrons and radiative Auger background (RAE) are used for this spectral deconvolution. The incident photon flux $\Phi_0(E_0)$ and the solid angle of detection $\frac{\Omega}{4\pi}$ are known due to the use of calibrated instrumentation \cite{Beckhoff2008}. The sample specific attenuation correction factor $M_{i,E0}$ for the incident ($E_0$) - as well as the fluorescence radiation ($E_i$) is calculated according to Eq. \ref{eq:M} using the experimentally determined attenuation coefficients $\mu_S(E_0)\rho d$ and $\mu_S(E_i) \rho d$.

\begin{figure}[htbp]
\centering
\includegraphics[width=0.78\textwidth]{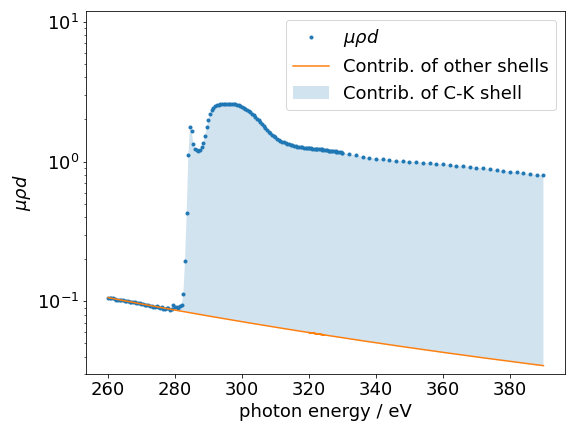}
	\caption{Experimental $\mu_S(E_0)\rho d$ (blue dots) for the employed carbon thin foil sample and its separation into the lower bound shells $\tau_{i,LM}(E_0)\rho d$ (orange) as well as $\tau_{K}(E_0)\rho d$ (light blue).}
\label{fig:tau}
\end{figure}

In addition, $\tau_K(E_0)\rho d$ as the sample specific subshell photoionization cross section for the K-edge of element of interest and at the chosen incident photon energies has to be known. The mass attenuation coefficient $\mu_S(E_0)$ is the sum of all shells’ photoionization probabilities $\tau_S(E_0)$ as well as of the cross sections for coherent $\sigma_{Coh}(E_0)$ and incoherent $\sigma_{Inc}(E_0)$ scattering. In the soft X-ray regime, the scattering contributions are negligible, thus $\mu_S(E_0) \approx \tau_S(E_0)$. To derive the sample specific K-shell ionization cross section of element $i$ $\tau_{K}(E_0)\rho d$ from the measured total ionization cross section $\tau_S(E_0)\rho d$, the latter must be separated into the various shells contributions \cite{M.Kolbe2012,P.Hoenicke2016a}. The lower bound shells of the element $\tau_{i,LM}(E_0)$ (orange) can be subtracted from $\tau_S(E_0)\rho d$ in order to derive $\tau_{K}(E_0)\rho d$ (blue shaded area). This is shown for the case of the carbon K-edge in Fig. \ref{fig:tau}.

\begin{figure}[htbp]
\centering
\includegraphics[width=0.78\textwidth]{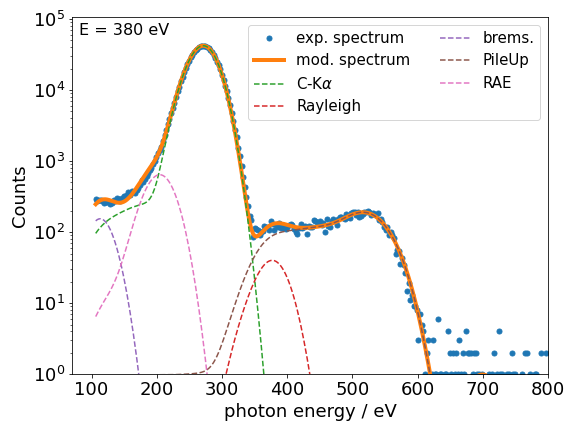}
	\caption{Experimental spectrum taken on the Carbon foil ($E_0$ = 380 eV) including the detector response function of the carbon K$\alpha$ fluorescence line (green dashed line), a pile-up contribution (brown dashes) as well as scattered incident radiation (red dashes). The radiative Auger emission (RAE) is shown as magenta dashed line and was modeled using a simple approximation based on literature spectra from highly resolving instruments \cite{Valjakka_1986, Anagnostopoulos_2018}.}
\label{fig:spek}
\end{figure}

The fluorescence intensities of the carbon K$\alpha$ lines are derived from the SDD spectra taken for each excitation energy by means of a spectral deconvolution, using the known detector response functions \cite{F.Scholze2009} for the relevant fluorescence lines. In addition, relevant background contributions for bremsstrahlung from photo electrons, resonant Raman scattering or radiative Auger emission can be taken into account if necessary. Due to the soft photon energies of the fluorescence lines, the automated pile-up rejection of the SDD system is not working properly and pile-up effects are visible in the spectra. These are also modeled using an empirically shaped object. The determined counts in the pile-up peak are later added to the derived fluorescence events for the fluorescence line causing the pile up events. In Fig. \ref{fig:spek}, a fluorescence spectrum as well as the respective modeled spectrum are shown for the carbon foil at an incident photon energy of $E_0 = 390$ eV. The derived count rates $D_{i,K}$  for the carbon K$\alpha$ lines are used to determine the detected fluorescence photon flux $\Phi_{i}^d (E_0)$ by normalization to the SDD’s detection efficiency for the respective fluorescence lines photon energy.

\subsection{\label{sec:level22}GIXRF-XRR determination}
\begin{figure}[htbp]
\centering
\includegraphics[width=0.9\textwidth]{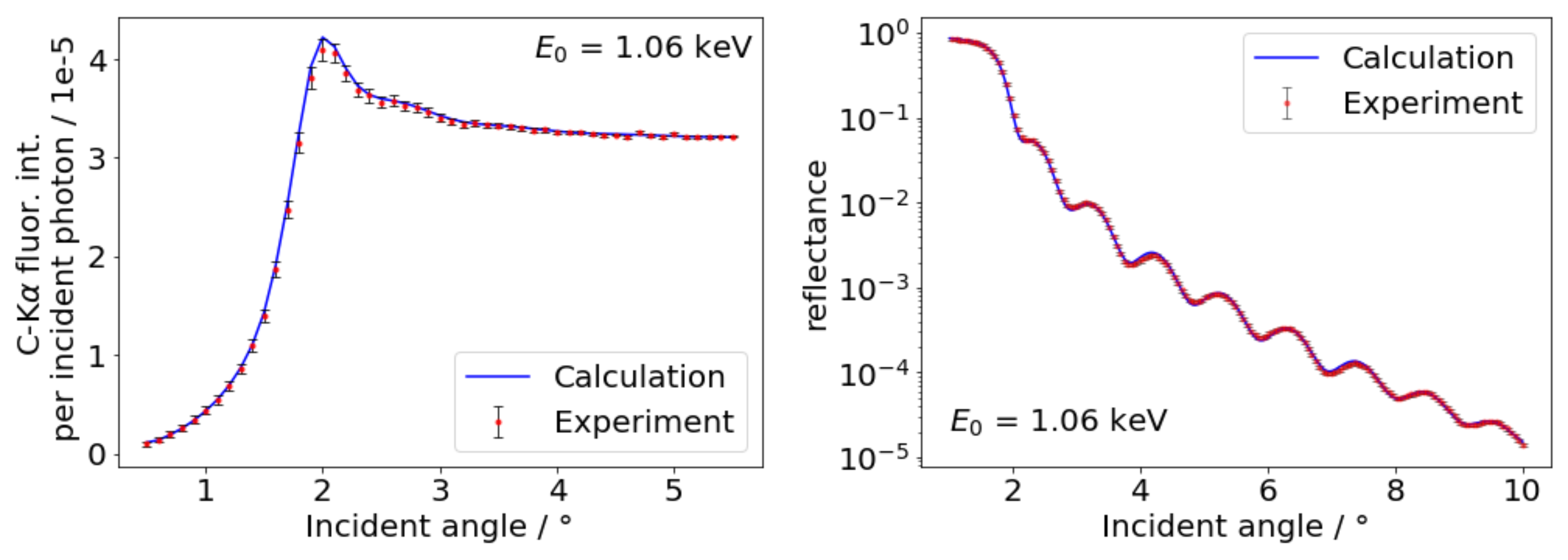}
	\caption{Comparison of the experimental reference-free GIXRF-XRR data with the model calculations. The GIXRF data is shown on the left, the XRR data on the right. See text for further details.}
\label{fig:GIXRF-XRR}
\end{figure}

In a previous work \cite{P.Hoenicke2016a}, it was already shown how a new experimental value for an atomic fundamental parameter can be validated employing XRR and XRF experiments with a thin film sample. Here, we go one step further and use reference-free GIXRF-XRR \cite{Hoenicke2019} experiments on a carbon thin film on silicon in order to determine the K-shell fluorescence production cross section (FPCS, product of fluorescence yield and photonionization cross section). For this purpose, a 30 nm thick carbon layer was coated on a silicon wafer piece (1 cm x 4 cm) using a dual ion beam deposition (DIBD) instrument designed for high precision nm- and sub-nm coating. Layers in DIBD typically grow in an amorphous state (a-C), as was the case here, too, rather than in crystalline form. However, the material density can be tuned in certain ranges by varying the running parameters of the two ion beams applied, from rather low density a-C to so called "diamond-like carbon" (d-C) with about 50\% higher density. The reference-free GIXRF-XRR were also performed at the PGM beamline \cite{F.Senf1998} employing an incident photon energy of 1.06 keV. The experimental data including a basic evaluation (spectra deconvolution, normalization to incident photon flux and solid angle of detection) are shown in Fig. \ref{fig:GIXRF-XRR}.

For the determination of the FPCS from the experimental data, a quantitative combined modeling is required. For this purpose, a model based on a contamination layer on carbon on native oxide covered silicon was used. For each layer, with the exception of the substrate, the thickness, relative density, and the roughness were used as  model parameters. Interfacial mixing was also considered. The modeling process is realized using the Sherman equation \cite{Sherman1955}, which is shown below. 

\begin{align}
\frac{4\pi\sin\theta_i}{\Omega(\theta_i)}\frac{F(\theta_i,E_i)}{\Phi_0\epsilon_{E_f}} &= W_i \rho \tau(E_i) \omega_k dz \cdot \sum_{z} P(z) \cdot I_{XSW}(\theta_i,E_i,z) \cdot \exp\left[-\rho\mu_{E_f}z\right]\textrm{.}
\label{eq:sherman}
\end{align}

Here, the experimentally derived fluorescence count rate $F(\theta_i,E_i)$ of C-K$\alpha$ radiation, excited using photons of energy $E_i$ at an incident angle $\theta_i$ is the essential measurand. A normalization on the effective solid angle of detection $\frac{\Omega(\theta_i)}{4\pi}$, the incident photon flux $\Phi_0$, and the detection efficiency of the used fluorescence detector $\epsilon_{E_f}$ is required. By calculating the X-ray standing wave field intensity distribution $I_{XSW}(\theta_i,E_i,z)$, a numerical integration in conjunction with the depth distribution $P(z)$ of the element of interest and an attenuation correction factor, the experimental data can be reproduced. For a quantitative modeling, the atomic fundamental parameters, namely the photo ionization cross section $\tau(E_i)$ and the fluorescence yield $\omega_k$, and material-dependent parameters, e.g. the weight fraction $W_i$ of element $i$ within the matrix as well as the density $\rho$ of the matrix must be considered. In the case of the carbon layer used here, $W_i$ is unity, and the density $\rho$ is made up of the product of bulk density and modeled relative density $\rho_{bulk} \rho_{rel}$. We used the bulk density of graphite ($2.26$ g/cm$^3$) as DIBD deposited amorphous carbon layer densities are usually in this regime \cite{Voevodin_1996}.

The relevant optical constants were taken from X-raylib \cite{T.Schoonjans2011} using $\rho_{bulk}$ and are also scaled using $\rho_{rel}$. The FPCS was also taken from X-raylib and is scaled employing a factor during the modeling. As only the product is relevant for the absolute value of the calculated fluorescence intensity, the fluorescence yield cannot be determined separately using this approach. The optimization was performed using a Markov chain Monte Carlo (MCMC) algorithm \cite{Foreman_Mackey_2013}.

The final model calculations are also shown in Fig. \ref{fig:GIXRF-XRR} and agree reasonably well with the experimental data. The determined layer thickness of the carbon layer is about 29 nm and thus well in line with the nominal value. For the scaling parameter of the FPCS we obtained a value of $0.990$ which means that the FPCS as calculated from X-raylib is already very accurate for the employed photon energy of 1.06 keV. Although this result is only valid for this photon energy, one can expect similar results for other energies not in the vicinity of the carbon K-shell attenuation egde. From existing literature data on the carbon mass attenuation coefficients \cite{REXADA_mu}, which where experimentally determined within the International initiative on X-ray fundamental parameters \cite{FPI}, one can assume that the X-raylib data for the mass attenuation coefficient of carbon at 1.06 keV is accurate with an estimated uncertainty of 5 \%. Thus, using the x-raylib data on the photoionization cross section for the carbon K-shell, the determined scaling applied to the X-raylib fluorescence yield for the carbon K-shell results in a value of $2.55 \times 10^{-3}$.

%
%

\section{\label{sec:level3}Relativistic calculations}

The MCDF method goes beyond the Coulomb approximation for the two-electron interaction by including the Breit interaction, which accounts for magnetic interactions and retardation effects in the calculation. It starts from the Dirac-Fock (DF) approximation, and takes into account the electronic correlation by minimizing an energy functional that is the mean value of the Hamiltonian within the virtual orbital space spanned by a selected number of configurations.  In this way, it is able to account for a larger amount of correlation with a smaller basis set than other methods \cite{Santos:2014ji}.
 
We used the relativistic General Purpose Multiconfiguration Dirac-Fock code (MCDFGME) developed by Desclaux \cite{92} and Indelicato \etal \cite{93} that has been improved consistently since its conception  \cite{277,266}. Quantum electrodynamics (QED) effects, such as self-energy, were included as perturbations \cite{812,278}.  

The wavefunctions and corresponding energies of the levels involved in all possible radiative and radiation-less transitions, were obtained using the optimum level (OL) method, considering full relaxation of both initial and final states, which provides accurate energies and wavefunctions. To deal with the orthogonality of the initial and final wavefunctions, the code uses the formalism described by L\"owdin \cite{285} in the calculation of radiative decay rates. 

For the calculation of the radiationless decay rates, the initial state wavefunctions were generated for configurations that contain one initial inner-shell vacancy while final state wavefunctions were generated for configurations that contain two higher shell vacancies. Continuum-state wavefunctions were obtained by solving the Dirac-Fock equations with the same atomic potential of the initial state, normalized to represent one ejected electron per unit energy.

The importance of electronic correlation in the calculated transition energies and probabilities has been shown by Santos \etal \cite{Santos:2006vf, Santos:2014ji}. In this work, due to the particular nature of the 2s$_{1/2}$ and 2p$_{1/2}$ electrons, correlation up to the 3d subshell was included. 

The K-shell fluorescence yield is defined as the relative probability that a K-shell vacancy is filled through a radiative transition considering both radiative and radiationless, or Auger, channels, i.e.,
\begin{equation}
\omega_{K}=\frac{\Gamma^{\mathrm{R}}}{\Gamma}=\frac{\Gamma^{\mathrm{R}}}{\Gamma^{\mathrm{R}}+\Gamma^{\mathrm{NR}}}
\label{FY}
\end{equation}
where $\Gamma, \Gamma^{\mathrm{R}}, \Gamma^{\mathrm{NR}}$ are the total, radiative, and radiation-less widths, respectively, of the initial hole level in the $K$ shell.
Further details may be found in  \cite{Sampaio:2014bm,Guerra:2018jo}.

%
%

\section{\label{sec:level4}Results and discussion}
\begin{figure}[htbp]
\centering
\includegraphics[width=0.78\textwidth]{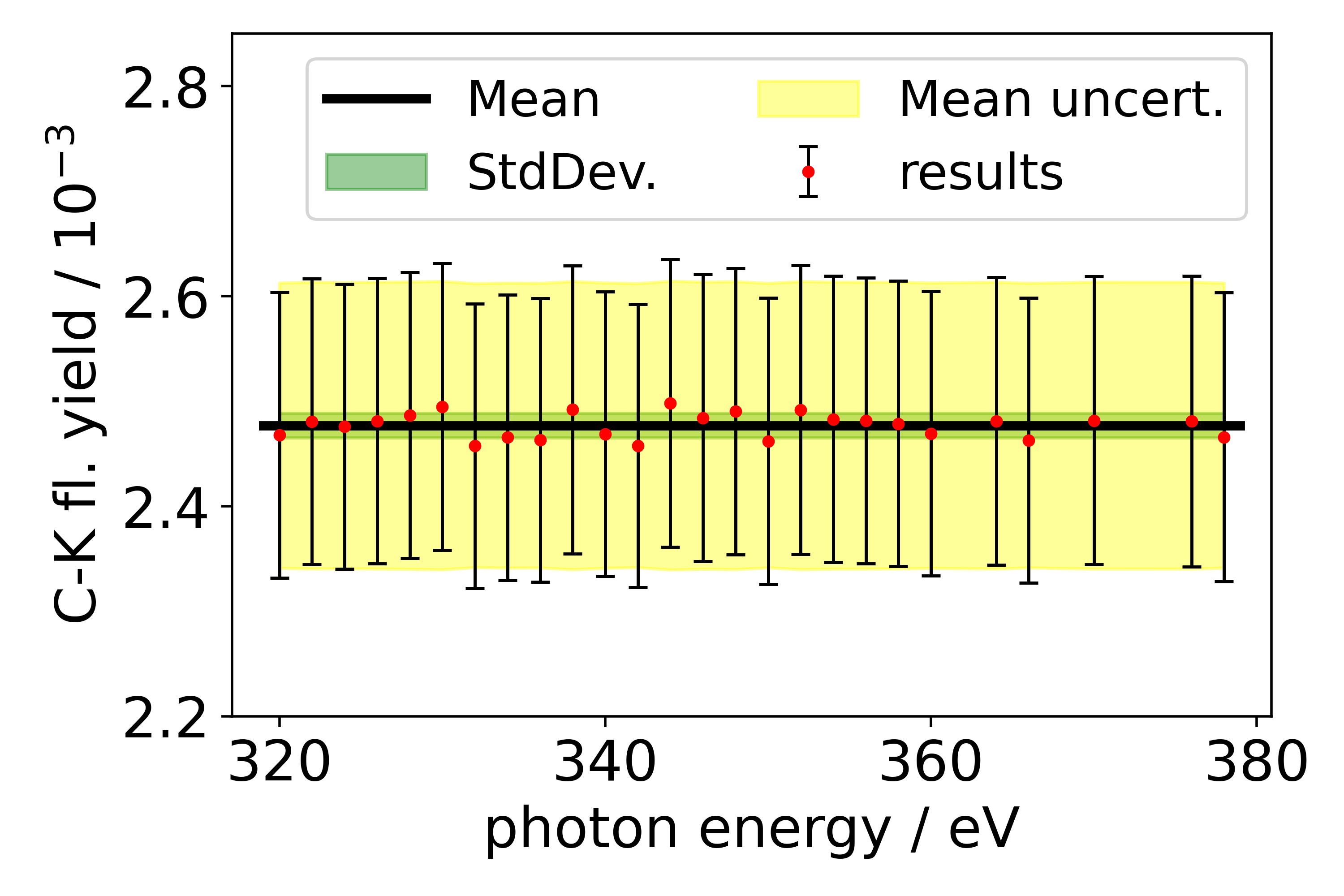}
	\caption{Determined carbon K-shell fluorescence yield values on the carbon foil for different incident photon energies.}
\label{fig:energycomp}
\end{figure}

The derived values for the sample specific photoionization cross sections $\tau_{i,K}(E_0)\rho d$ and the derived fluorescence photon fluxes $\Phi_{i}^d (E_0)$ can be used to calculate the K-shell fluorescence yields for the studied thin foil samples according to equation \ref{eq:prodCS}. This was done for each excitation energy and the results obtained on the carbon thin foil are shown in Fig. \ref{fig:energycomp}. As expected, the determined fluorescence yield does not show any significant dependence of the photon energy. This serves as an internal control for the derived photoionization cross sections $\tau_{i,K}(E_0)\rho d$. The shown errorbars represent the achieved uncertainty at each photon energy, which slightly changes due to varying counting statistics and the varying spectral deconvolution accuracy. From the shown results, we have calculated mean values for both the carbon K-shell fluorescence yield (shown as a black horizontal line in Fig. \ref{fig:energycomp}) and its experimental uncertainty and performed the same analysis for the other thin foil samples. 

The uncertainty budget of the presented results is calculated on the basis of the relative uncertainty contributions of the relevant parameters. The main contributors to the total uncertainty budget are the determined photo ionization cross sections (~2.5 \%), the attenuation correction factors (2 \%) and the detection efficiency of the employed silicon drift detector (3 \%). The latter is higher in the case of carbon fluorescence as the synchrotron beamline used for the calibration of the SDD suffers from carbon contamination of the optics resulting in a degraded spectra purity. The uncertainty budget one can achieve by employing PTB's reference-free XRF approach towards the determination of atomic fundamental parameters is discussed in more detail in ref. \cite{Unterumsberger2018}.

The  calculated value using the MCDF method, including relativistic and QED corrections, was determined to be $\omega_{K}=2.67\times10^{-3}$. Here, an estimated uncertainty of 10\% is obtained by error propagation of Eq. (\ref{FY}). 
The individual uncertainty of $\Gamma^{\mathrm{R}}$ was obtained as the average of the differences in transition rates between the length and the velocity gauge, weighted by the transition rates themselves. Due to the impossibility of using the same approach to the radiationless rates, as the  quality of the wave functions should be similar for two-hole states,  we have taken   the uncertainty of $\Gamma^{\mathrm{NR}}$  as the same as the uncertainty of  $\Gamma^{\mathrm{R}}$ .

\begin{figure}[htbp]
\centering
\includegraphics[width=0.78\textwidth]{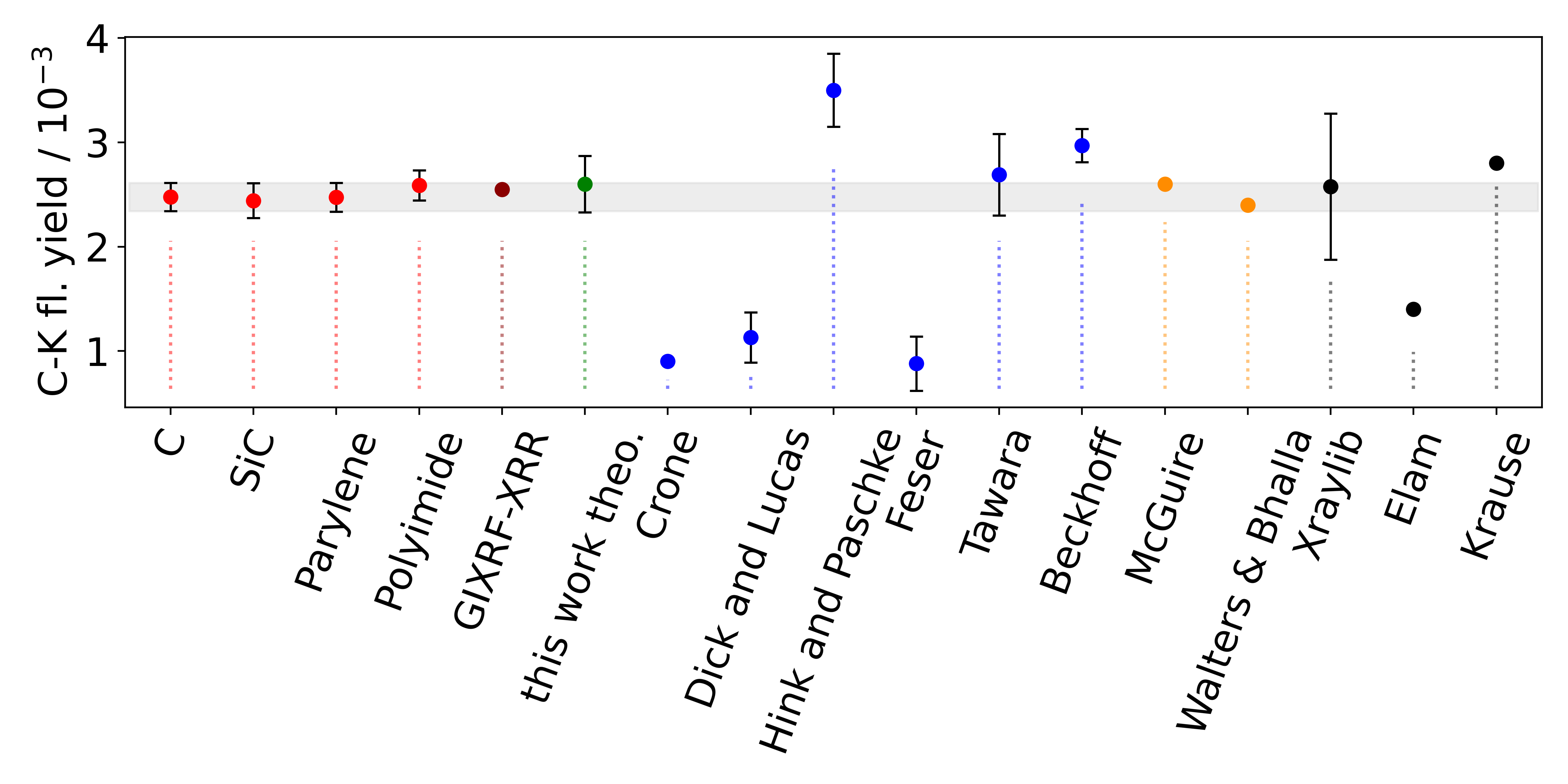}
	\caption{Comparison of the here determined experimental and theoretical K-shell fluorescence yield values for carbon and available literature values as listed in tab. \ref{tab:addlabel}.}
\label{fig:litcomp}
\end{figure}

In Table~\ref{tab:addlabel} as well as fig. \ref{fig:litcomp}, the K-shell fluorescence yield values for carbon (experimental and theoretical) obtained in this work are listed together with the available experimental, theoretical, and semi-empirical literature values. 
We find a very good agreement between our experimental and theoretical results, since all our experimental values lie within the theoretical error bars, which mutually validates our results.This agreement is in the order of our previous works on other fluorescence yields for soft X-ray attenuation edges \cite{909,Guerra:2018jo}. The slight discrepancy may arise from condensed matter effects, not taken into account in our calculations, due to the fact that the experimental values were obtained using free-standing thin foils of elemental carbon, polyimide, parylene and silicon carbide, and not using isolated carbon atoms. 

When comparing with the available literature data, a good agreement can be found with the value determined by Tawara \etal \cite{H.Tawara1973}, McGuire \cite{McGuire1970} and the Walters and Bhalla \cite{D.L.Walters1971} value. The agreement with respect to the value \cite{B.Beckhoff2001a} published earlier from our group is not satisfying enough. But, the older experiments were performed with a thin-window liquid nitrogen cooled Si(Li) detector having both poorer energy resolution and count-rate capabilities than a current SDD. As a result, the contributions shown in Fig. \ref{fig:spek} could not be deconvoluted and are thus all added to the carbon fluorescence increasing the obtained yield value. Comparing to the widely used value from Krause's tables \cite{Krause1979}, the agreement is worse as the Krause value is larger, but there is agreement if we consider his estimated uncertainty of at least 25 \%. The newest available fundamental parameter database X-raylib \cite{T.Schoonjans2011} provides a much better agreeing carbon K-shell fluorescence yield value as compared to Krause and by far considering Elam \etal \cite{W.T.Elam2002}.

\begin{table}[htbp]
\caption{K-shell fluorescence yield values ($\omega_\textrm{K} / 10^{-3}$) for C. The most commonly used sources are Elam \etal~\cite{W.T.Elam2002}, Krause~\cite{Krause1979} and X-raylib~\cite{T.Schoonjans2011} values due to their availability as consistent database.}
\label{tab:addlabel}
\begin{ruledtabular}
\begin{tabular}{llcc}
&  \multicolumn{1}{c}{$\textrm{Expt.}$}  &   \multicolumn{1}{c}{$\textrm{Theo.}$}  &  \multicolumn{1}{c}{$\textrm{Semi-emp.}$}  \\
\cline{1-4}
\\%
This work - Carbon        &		2.47(14) &	2.67(27)  & \\
This work - Polyimide	    &		2.59(15) &	  & \\
This work - Parylene	      &		2.47(14) &	  & \\
This work - Silicon Carbide&		2.47(17) &	  & \\
This work - GIXRF-XRR      & 2.55 & & \\
\\
Crone (1936) \cite{Crone1936}& 0.9 \\
Dick and Lucas (1970) \cite{Dick_1970}& 1.13(24) \\
Hink and Paschke (1971) \cite{Hink_1971}& 3.50(35) \\
Feser (1972) \cite{Feser_1972}& 0.88(26) \\
Tawara (1973) \cite{H.Tawara1973}& 2.69(39) \\
Beckhoff (2001) \etal\cite{B.Beckhoff2001a}& 2.97(16) \\
&  \\
McGuire (1970) \cite{McGuire1970} & & 2.60 \\
Walters and Bhalla (1971) \cite{D.L.Walters1971} & & 2.40 \\
&  \\
Krause (1979) \cite{Krause1979} & & & 2.8(7) \\
Elam \etal (2002)  \cite{W.T.Elam2002}& & & 1.40 \\
X-raylib (2011) \cite{T.Schoonjans2011} & & & 2.58 \\
\end{tabular}
\end{ruledtabular}
\end{table}

The uncertainties of the literature values, if provided, are all larger than the determined total uncertainty budget of the experimental data presented here. The largest difference can be found in comparison to the estimated uncertainty of Krause \cite{Krause1979}. Thus, the present reduction of the uncertainty of the C K-edge fluorescence yield allows to significantly improve the accuracy and to reduce the uncertainty of fundamental parameter based quantification results.

%
%

\section{\label{sec:level5}Conclusion}
The fluorescence yield for the carbon K shell has been experimentally determined using the reference-free XRF setup of PTB and free-standing thin foils for various carbon containing materials. In addition, a novel GIXRF-XRR based approach for fundamental parameter validation or determination is presented and furthermore, the K-shell yield has been theoretically calculated using the MCDF method, including relativistic and QED corrections. A very good agreement between the different results obtained in this work could be achieved. More importantly, the uncertainties of the carbon K-shell yield value could be significantly reduced as compared to existing literature data having a direct impact on the uncertainty of fundamental parameter based XRF quantification results. This is in particular important when such XRF methods are required as reference-material based XRF cannot be performed due to missing adequate reference materials.
%
%

\section{\label{sec:level6}Acknowledgements}
This project has received funding from the ECSEL Joint Undertaking (JU) under grant agreement No 875999 - IT2. The JU receives support from the European Union’s Horizon 2020 research and innovation
programme and Netherlands, Belgium, Germany, France, Austria, Hungary, United Kingdom, Romania, Israel. Parts of this research was performed within the EMPIR projects Aeromet II. The financial support of the EMPIR program is gratefully acknowledged. It is jointly funded by the European Metrology Programme for Innovation and Research (EMPIR) and participating countries within the European Association of National Metrology Institutes (EURAMET) and the European Union. This research was supported in part by FCT (Portugal) under research center grants UID/FIS/04559/2020 (LIBPhys) and by the project PTDC/FIS-AQM/31969/2017, "Ultra-high-accuracy X-ray spectroscopy of transition metal oxides and rare earths".


\bibliographystyle{model1-num-names}
\bibliography{lit}

\end{document}